\documentclass[12pt,preprint]{aastex}
\usepackage{apjfonts}

\lefthead{JUNG ET AL.}
\righthead{ }

\begin{document}
\title{KMT-2016-BLG-1820 and KMT-2016-BLG-2142: 
Two Microlensing Binaries Composed of Planetary-mass Companions and Very-Low-Mass Primaries
}


\author{
Youn~Kil~Jung$^{1}$, 
Kyu-Ha~Hwang$^{1}$,
Yoon-Hyun~Ryu$^{1}$, 
Andrew~Gould$^{1,2,3}$,
Cheongho~Han$^{4}$,
Jennifer~C.~Yee$^{5}$,
Michael~D.~Albrow$^{6}$, Sun-Ju~Chung$^{1,7}$, In-Gu~Shin$^{5}$, Yossi~Shvartzvald$^{8}$, Weicheng~Zang$^{9,10}$, 
Sang-Mok~Cha$^{1,11}$, Dong-Jin~Kim$^{1}$, Hyoun-Woo~Kim$^{1}$, Seung-Lee~Kim$^{1,7}$, 
Chung-Uk~Lee$^{1,7}$, Dong-Joo~Lee$^{1}$, Yongseok~Lee$^{1,11}$, Byeong-Gon~Park$^{1,7}$, Richard~W.~Pogge$^{2}$
}

\bigskip\bigskip
\affil{$^{1}$Korea Astronomy and Space Science Institute, Daejon 34055, Republic of Korea}
\affil{$^{2}$Department of Astronomy, Ohio State University, 140 W. 18th Ave., Columbus, OH 43210, USA}
\affil{$^{3}$Max-Planck-Institute for Astronomy, K$\rm \ddot{o}$nigstuhl 17, 69117 Heidelberg, Germany}
\affil{$^{4}$Department of Physics, Chungbuk National University, Cheongju 28644, Republic of Korea}
\affil{$^{5}$Harvard-Smithsonian Center for Astrophysics, 60 Garden St., Cambridge, MA 02138, USA}
\affil{$^{6}$University of Canterbury, Department of Physics and Astronomy, Private Bag 4800, Christchurch 8020, New Zealand}
\affil{$^{7}$Korea University of Science and Technology, 217 Gajeong-ro, Yuseong-gu, Daejeon 34113, Korea}
\affil{$^{8}$IPAC, Mail Code 100-22, Caltech, 1200 E. California Blvd., Pasadena, CA 91125, USA}
\affil{$^{9}$Physics Department and Tsinghua Centre for Astrophysics, Tsinghua University, Beijing 100084, China}
\affil{$^{10}$Department of Physics, Zhejiang University, Hangzhou, 310058, China}
\affil{$^{11}$School of Space Research, Kyung Hee University, Yongin 17104, Republic of Korea}

\begin{abstract}
We present the analyses of two short-timescale $(t_{\rm E} \sim 5~{\rm days})$ 
microlensing events, KMT-2016-BLG-1820 and KMT-2016-BLG-2142. In both light curves, 
the brief anomalies were clearly captured and densely covered by the Korea Microlensing 
Telescope Network survey. From these analyses, we find that both events have small 
Einstein radii of $\theta_{\rm E} = 0.12~{\rm mas}$, suggesting that the binary-lens 
systems are composed of very low-mass components and/or are located much closer to the 
lensed stars than to Earth. From Bayesian analyses, we find that these binaries 
have total system masses of $0.043_{-0.018}^{+0.043}~M_{\odot}$ and 
$0.088_{-0.041}^{+0.120}~M_{\odot}$, implying that they are well within the 
very-low-mass regime. The estimated lens-component masses indicate that the binary lenses 
consist of a giant-planet/brown-dwarf pair (KMT-2016-BLG-1820), and a dark/faint object pair 
(KMT-2016-BLG-2140) that are located near the deuterium-burning and hydrogen-burning mass 
limits, respectively. Both lens systems are likely to be in the Galactic disk with estimated 
distances of about $6$ kpc and $7$ kpc. The projected lens-components separations are $1.1$ AU 
and $0.8$ AU, and the mass ratios are $0.11$ and $0.20$. These prove that the microlensing 
method is effective to identify these closely-separated very-low-mass binaries 
having low mass-ratios. 
\end{abstract}

\section{Introduction}

Over the 20 years following the first discoveries by \citet{rebolo95} and \citet{nakajima95}, 
thousands of very low-mass (VLM; $M \leq 0.2 M_{\odot}$) stellar and substellar objects 
have been discovered\footnote{http://www.astro.umontreal.ca/$\sim$gagne/listLTYs.php}. 
However, despite the large assemblage of these cool objects, 
their formation still remains as an open question. Various scenarios have been 
proposed to explain their origins, including ejection from multiple prestellar cores 
(e.g., \citealt{reipurth01}), turbulent fragmentation of gas in protostellar clouds 
(e.g., \citealt{padoan04}), photoionizing radiation from massive nearby stars 
(e.g., \citealt{whitworth04}), and fragmentation of unstable prestellar disks 
(e.g., \citealt{stamatellos09}).

Observational studies of VLM binaries can provide effective diagnostics for testing 
the VLM formation scenarios. This is because the formation mechanisms leave their 
own traces on the statistical properties of binaries such as frequency, orbit 
separation, and mass-ratio distributions (e.g., \citealt{bate09}). In addition, 
binaries can provide a model-independent way to determine physical properties 
including masses. For the comprehensive study of the formation scenarios, 
hence, it is essential to obtain unbiased VLM binary samples from various 
detection methods that are effective with respect to different population of VLM objects.

Despite its importance, the sample of VLM binaries still remains incomplete. 
Up to now, the most productive observational method to detect VLM binaries 
is direct imaging (e.g., \citealt{close07}). Due to its current limit of 
angular resolution, however, it is difficult to identify close 
$(\lesssim 3~{\rm AU})$ binaries by this method, and thus the samples 
could be biased toward wide binaries (e.g., \citealt{burgasser07}). 
For the same reason, the method is sensitive to roughly equal mass ratio 
$(q = M_{2}/M_{1} \gtrsim 0.5)$ binaries confined to the solar neighborhood 
and to star forming regions. Moreover, it is difficult to precisely determine 
the masses of these objects from spectra because of their faintness and 
long orbital period 
\footnote{We note that VLM binaries discovered from blended-light spectra 
(e.g., \citealt{bardalez14}) and from astrometric perturbations 
(e.g., \citealt{sahlmann13}) have little separation bias. In addition, the 
astrometric perturbation method is also sensitive to unequal mass ratio binaries.}.

Microlensing can complement the direct imaging method by detecting binaries 
that are difficult to identify by other methods. The lensing phenomenon only 
relies on the gravitational field of a lensing object and thus enables one 
to detect binaries composed of very faint and even dark objects with 
masses down to brown dwarf (BD) and planetary regimes. Because 
microlensing events arise on Galactic scales, they can probe binaries located 
far beyond the solar neighborhood. In addition, the method can detect even 
tight $(\leq 1~{\rm AU})$ binaries because of its high sensitivity to binaries
with small separations. With these advantages, the microlensing technique has 
discovered various types of VLM objects such as isolated BDs 
(e.g., \citealt{gould09,zhu16, chung17}), a BD hosting a planetary companion 
(e.g., \citealt{han13,jung18}), BD-BD binaries (e.g., \citealt{choi13}), 
and BDs around VLM stars (e.g., \citealt{jung15,han17}).

In this work, we report two binary systems that are composed of planetary-mass 
companions and very low-mass primaries. They were discovered from the analysis 
of short timescale events found from the Korea Microlensing Telescope Network 
\citep[KMTNet:][]{kim16} survey conducted in 2016 season.

\section{Observations}

These VLM binaries were discovered in lensing events KMT-2016-BLG-1820 
and KMT-2016-BLG-2142. The equatorial and galactic coordinates of the events are 
$({\rm RA}, {\rm Dec})_{\rm J2000} = $(17:55:03.54, $-29$:31:00.91) 
$[(l,b) = (0.56, -2.07)]$ and 
$({\rm RA}, {\rm Dec})_{\rm J2000} = $(17:52:26.88, $-29$:23:04.42) 
$[(l,b) = (0.38, -1.51)]$, respectively.

The events were detected from the KMTNet survey. The survey was conducted 
toward the Galactic bulge field using three identical 1.6 m telescopes that
are globally distributed at the Cerro Tololo Inter-American Observatory in 
Chile (KMTC), South African Astronomical Observatory in South Africa (KMTS), 
and Siding Spring Observatory in Australia (KMTA). Both events are located in 
two overlapping fields (KMTNet BLG02 and BLG42), resulting in 
a combined $4~{\rm hr}^{-1}$ cadence. Most KMTNet images were obtained in 
$I$-band and some $V$-band images were taken to measure the colors of the 
lensed stars. For the analysis, the data of the individual events were 
reduced using the pySIS package \citep{albrow09}, and their errors were 
re-scaled using the method of \citet{yee12}.

In the summer of 2017, KMT-2016-BLG-1820 and KMT-2016-BLG-2142 were initially 
identified as ``clear'' and ``possible'' microlensing events through a human 
review of lensing candidates found by the KMTNet Event Finder algorithm \citep{kim18a}. 
The light curves used for the algorithm were constructed based on the Difference 
Imaging Analysis method \citep[DIA:][]{alard98,wozniak00}, for which catalog stars 
are mainly assembled from the OGLE-III star catalog \citep{szymanski11}. During 
this inspection, KMT-2016-BLG-1820 and KMT-2016-BLG-2142 were discovered 
on the light curves of $I = 19.4$ and $I = 19.8$ catalog stars, respectively. 
In contrast to KMT-2016-BLG-1820, KMT-2016-BLG-2142 was only rated as ``possible'' 
(rather than ``clear'') microlensing at this stage due to its relatively noisy DIA 
light curve. In the course of carrying out the automated pySIS reductions 
in preparation for the data release in the winter of 2017 \citep{kim18b}, however, 
it was recognized that the high noise in the DIA light curve originated from a position 
offset between the true lensed star and the catalog star. From the pySIS light curve, 
the lensing anomaly of KMT-2016-BLG-2142 was then clearly identified.

\section{Analysis}

\begin{figure}[th]
\epsscale{0.9}
\plotone{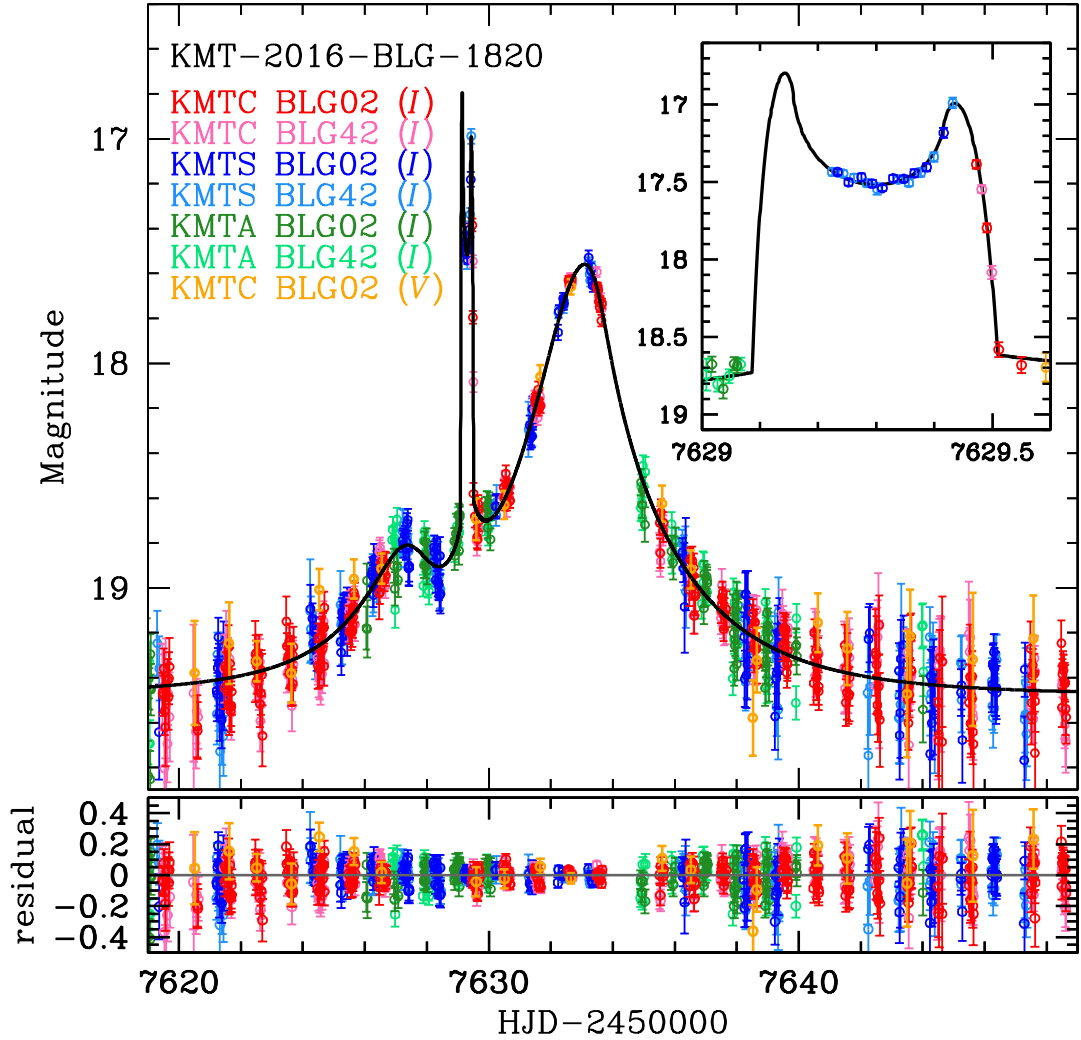}
\caption{\label{fig:one}
Light curve of KMT-2016-BLG-1820. 
The right inset shows the caustic crossing centered at ${\rm HJD}' \sim 7629.3$. 
The black curve on the data is the best-fit model. 
The residuals from the model are presented in the lower panel. 
Note that the V-band data are only used for the source color measurement.
}
\end{figure}

\begin{figure}[th]
\epsscale{0.9}
\plotone{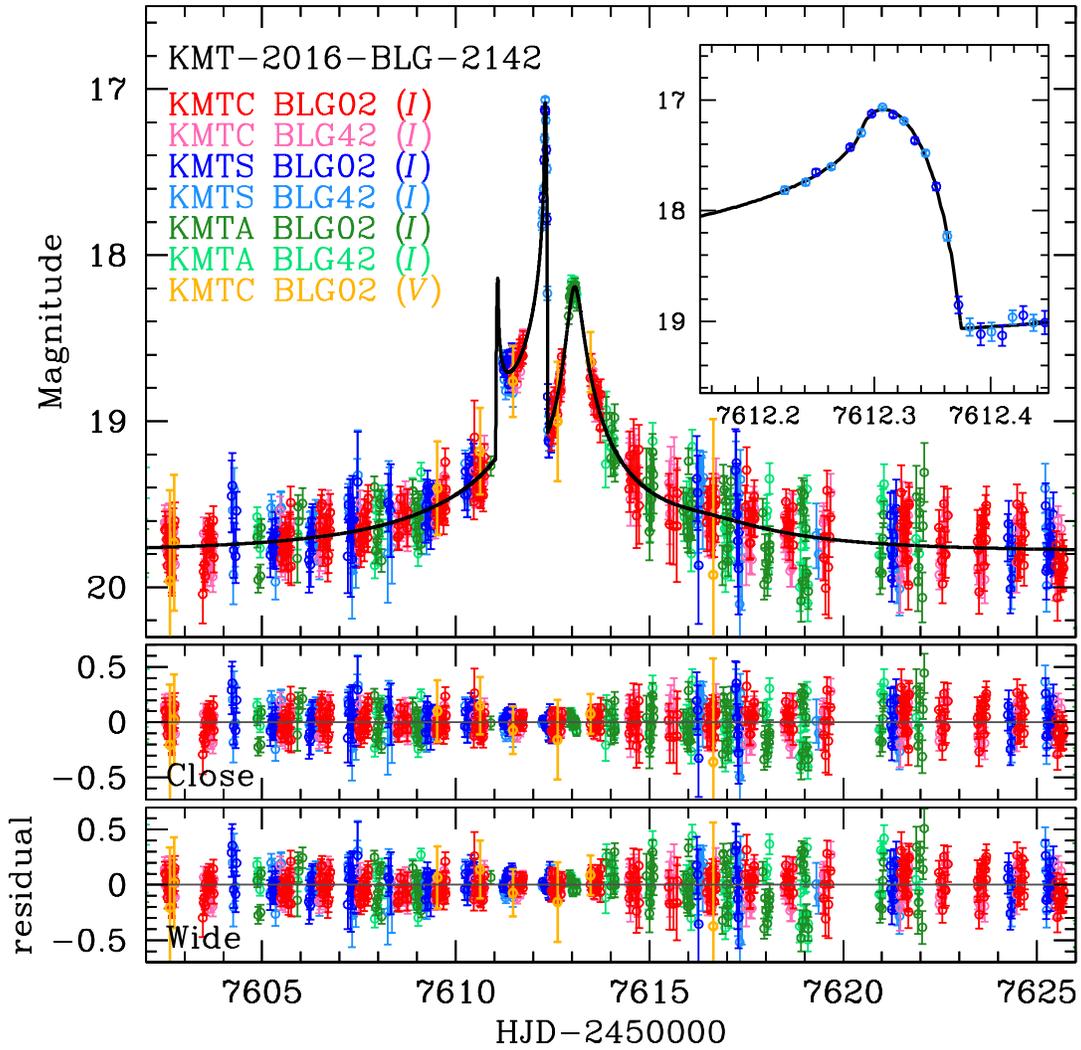}
\caption{\label{fig:two}
Light curve of KMT-2016-BLG-2142. 
The right inset shows the caustic exit centered at ${\rm HJD}' \sim 7612.3$. 
The black curve on the data is the best-fit ``Close'' model. 
The lower two panels show the residuals from the ``Close'' and ``Wide'' solutions.
}
\end{figure}

Figure~\ref{fig:one} and~\ref{fig:two} show the light curves of 
KMT-2016-BLG-1820 and KMT-2016-BLG-2142, respectively. Both light 
curves exhibit strong spikes with U-shaped troughs. Also, there 
exists a weak bump before the spike near 
${\rm HJD}'(={\rm HJD} - 2450000~{\rm days}) \sim 7627$ 
for KMT-2016-BLG-1820 and a strong bump after the spike near 
${\rm HJD}' \sim 7613$ for KMT-2016-BLG-2142. These spikes and bumps 
are typical binary-lensing anomalies that are generally produced by 
caustic-crossings and caustic-approaches of lensed stars, respectively. 
Thus, we investigate the events based on the binary-lens interpretation.

\begin{figure*}[th]
\epsscale{0.9}
\plotone{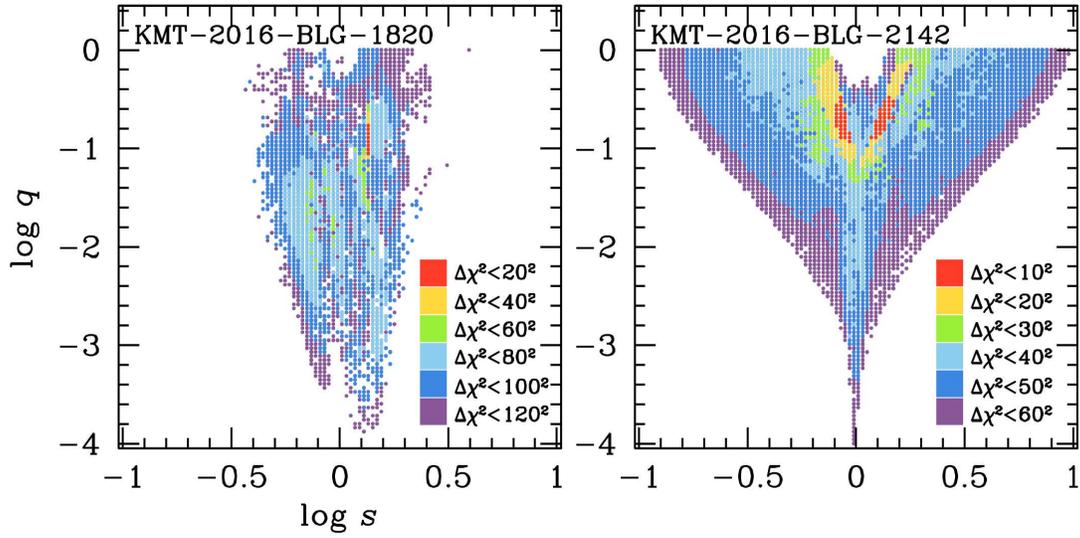}
\caption{\label{fig:three}
$\Delta\chi^{2}$ distributions in the $({\rm log}~s, {\rm log}~q)$ space 
for KMT-2016-BLG-1820 (left panel) and KMT-2016-BLG-2142 (right panel). 
In both panels, the spaces are divided by $(100,100)$ fixed grids and 
the inspected ranges are $-1 < {\rm log}~s < 1$ and $-4 < {\rm log}~q < 0$, respectively.
Note that $\Delta\chi^2$ contours are differently color coded for each event. 
}
\end{figure*}

To analyze to the light curve, we use the parametrization  
and follow the modeling procedure discussed in \citet{jung15}. 
We initially conduct a search over a grid of $(s, q, \alpha)$, where $s$ 
is the projected lens-component separation and $\alpha$ is the trajectory 
angle. We note that all of the lengths are scaled to the angular Einstein 
radius of the lens, $\theta_{\rm E}$. From this initial search, we explore 
local $\chi^{2}$ minima on the grid-parameter space. Figure~\ref{fig:three} 
shows our derived $\Delta\chi^2$ surface over the $({\rm log}~s, {\rm log}~q)$ 
plane. For KMT-2016-BLG-1820, we find only one local minimum. 
For KMT-2016-BLG-2142, we identify a pair of local solutions 
(marked as ``Close'' and ``Wide'') for which the mass ratios 
are similar but the separations have opposite signs of ${\rm log}~s$, 
i.e., a close/wide degeneracy \citep{griest98,dominik99}. To find the 
global minimum, we then seed these local solutions into new MCMCs 
and allow all parameters to vary. Note that, in this last stage, 
we additionally explore the parameter space including the microlens parallax 
\citep{gould92,gould04} and orbital motion of the lens \citep{dominik98,jung13}, 
but we cannot measure the signals because of the faintness and 
short timescales of the events.

\begin{figure}[th]
\epsscale{0.8}
\plotone{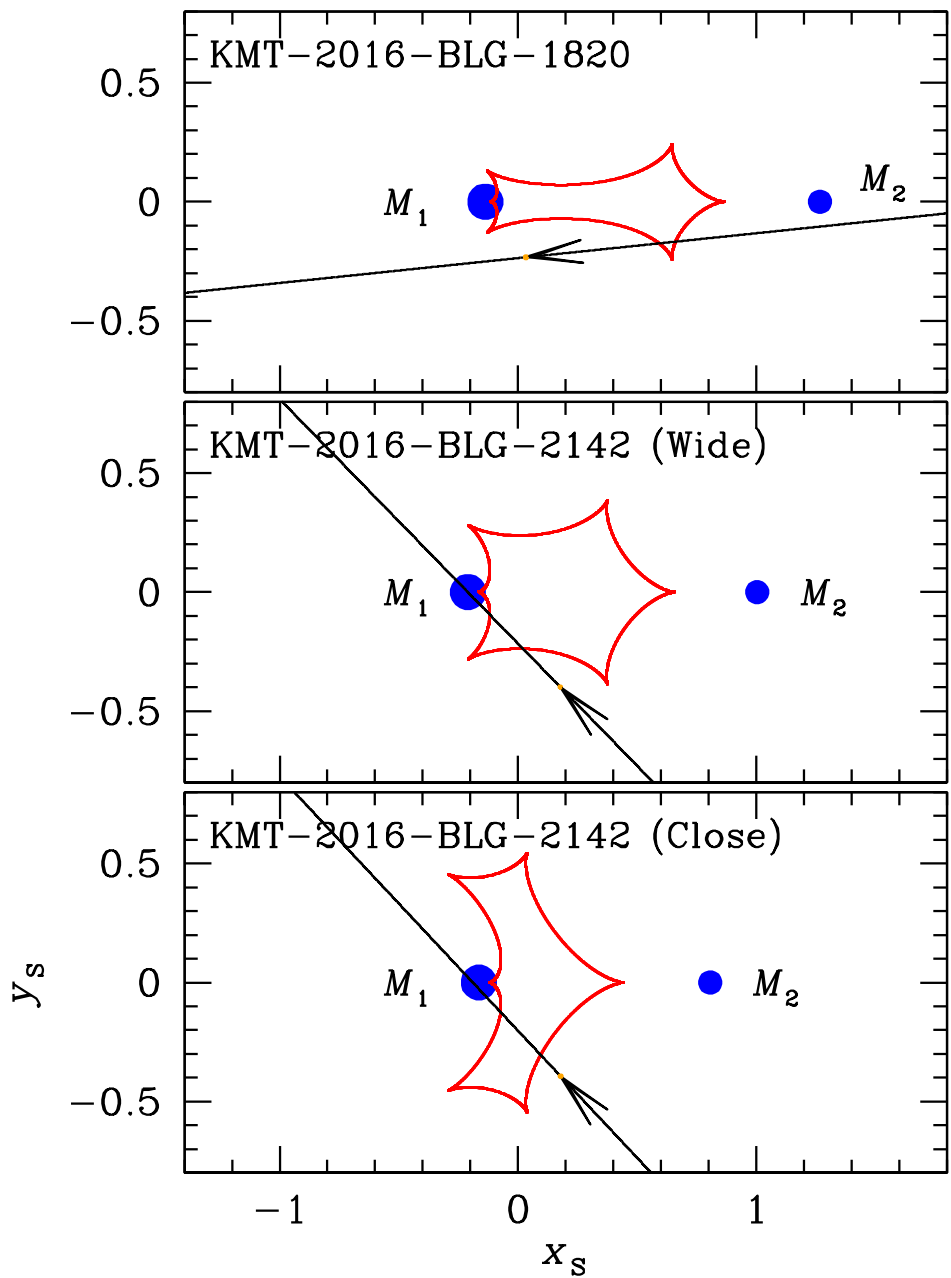}
\caption{\label{fig:four}
Caustic geometries of KMT-2016-BLG-1820 (upper panel) and 
KMT-2016-BLG-2142 (middle and lower panels). In each panel, 
the caustic is represented by the red closed curve. The straight 
line shows the source trajectory and the arrow on the line indicates 
the direction of source motion. The two blue circles are the 
position of primary ($M_{1}$) and secondary ($M_{2}$) masses. 
}
\end{figure}

Table~\ref{table:one} gives the best-fit solutions of the individual events. 
The corresponding model curves are presented in Figure~\ref{fig:one} and~\ref{fig:two}. 
Also presented in Figure~\ref{fig:four} are the configurations of the lens systems 
where the source trajectories (straight lines with arrows) with respect to the caustics 
(red closed curves) and the lens components (two blue circles) are shown. 
We find that both events have similar lensing characteristics in the sense that 
the derived Einstein timescales $t_{\rm E}$ are relatively short 
$(t_{\rm E} \lesssim 5~{\rm days})$ and mass ratios are relatively low $(q \lesssim 0.2)$. 
Considering that $t_{\rm E}$ is proportional to the lens mass 
($\propto \sqrt{M_{\rm tot}}$), these give a clue that the secondary masses could  
correspond to those of substellar objects. For each event, we find that the U-shape 
variation was generated by the source crossing over the resonant caustic that 
forms when the projected separation is similar to $\theta_{\rm E}$. The weak bump 
in KMT-2016-BLG-1820 near ${\rm HJD}' \sim 7627$ was generated when the source 
approached one of the cusps located close to the lower-mass lens component, while the strong 
bump in KMT-2016-BLG-2142 near ${\rm HJD}' \sim 7613$ was produced when the source approached 
one of the central cusps located close to the higher-mass component. For KMT-2016-BLG-2142, 
we find that the ``Close'' model is favored over the ``Wide'' model by 
$\Delta\chi^{2} \sim 41 ~ ( > 6\sigma)$, which is statistically high enough to exclude 
the wide-binary interpretation. Hence, we only consider the ``Close'' solution.

\begin{figure*}[th]
\epsscale{0.9}
\plotone{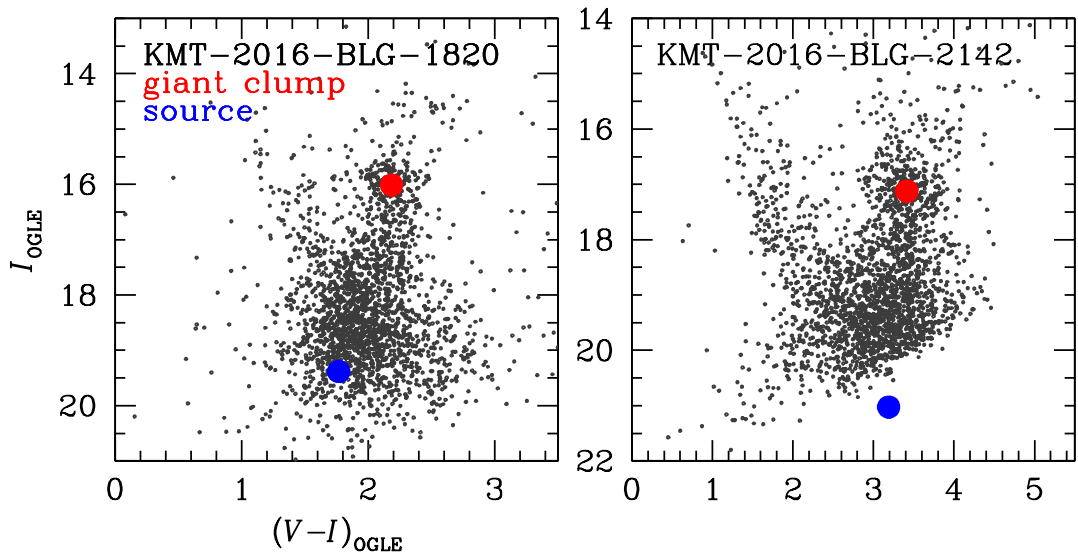}
\caption{\label{fig:five}
Instrumental color-magnitude diagrams of KMT-2016-BLG-1820 
(left panel) and KMT-2016-BLG-2142 (right panel), calibrated to 
OGLE-III photometry map \citep{szymanski11}. In each panel, the red and 
blue dots indicate the positions of giant clump centroid and source, 
respectively. 
}
\end{figure*}

For both events, the finite source effects are clearly detected from which 
we can measure the normalized source radius $\rho_{*}$. These enable us 
to determine $\theta_{\rm E}$ and the lens-source relative proper 
motion $\mu$ by
\begin{equation}
\theta_{\rm E} = {\theta_{*} \over \rho_{*}};~~~~~ \mu = {\theta_{\rm E} \over t_{\rm E}},   
\label{eq1}
\end{equation}
where $\theta_{*}$ is the angular radius of the source. To determine $\theta_{*}$, 
we adopt the standard method of \citet{yoo04}. First, we build the KMTNet star 
catalog using the pyDIA reductions and calibrate the brightness of stars using 
the OGLE-III catalog \citep{szymanski11}. Second, we estimate the source $(V-I, I)$ from the model 
and then place the source star on the constructed color-magnitude diagram (CMD). Third, we measure the 
relative source position using the giant clump (GC) centroid as a reference, 
i.e., $\Delta (V-I, I)$. Fourth, we estimate the de-reddened source position 
$(V-I, I)_{0}$ as   
\begin{equation}
(V-I, I)_{0} = \Delta (V-I, I) + (V-I, I)_{0,\rm GC}.
\label{eq2}
\end{equation}
Here the de-reddened GC centroid $(V-I, I)_{0,\rm GC}$ are known from independent 
observations \citep{bensby13,nataf13}. Finally, we deduce $\theta_{*}$ by first 
converting $(V-I)_{0}$ to $(V-K)_{0}$ from the color-color relation 
\citep{bessell88} and then by applying the $(V-K)_{0} - \theta_{*}$ relation 
of \citet{kervella04}. In Table~\ref{table:two}, we summarize our derived offsets 
$\Delta (V-I, I)$, de-reddened GC and source positions, angular source radii, 
angular Einstein radii, and relative lens-source proper motions of the individual 
events. In Figure~\ref{fig:five}, we present the GC and source positions of the 
individual events in the corresponding CMDs.

The angular Einstein radius is connected to $M_{\rm tot}$ and 
the lens distance $D_{\rm L}$ by 
\begin{equation}
\theta_{\rm E} \equiv \sqrt{{\kappa}M_{\rm tot}\pi_{\rm rel}};~~~~~\pi_{\rm rel} = {\rm AU}\left({1 \over D_{\rm L}} - {1 \over D_{\rm S}}\right),
\label{eq3}
\end{equation}
where $\kappa = 4G/(c^{2}{\rm AU}) \sim 8.14~{\rm mas}/M_{\odot}$, 
$\pi_{\rm rel}$ is the lens-source relative parallax, 
and $D_{\rm S}$ denotes the distance to the source, 
which is $D_{\rm S} \sim 8~{\rm kpc}$ for a typical bulge star. 
For a lensing event caused by a low-mass stellar lens located 
halfway between the observer and a bulge source, 
the size of $\theta_{\rm E}$ is then 
\begin{equation}
\theta_{\rm E} \sim 0.5~{\rm mas} \left({M_{\rm tot} \over 0.3~M_{\odot}}\right)^{1/2} \left({\pi_{\rm rel} \over 0.12~{\rm mas}}\right)^{1/2}. 
\label{eq4}
\end{equation}
For our analyzed events, the measured angular Einstein radii, 
which are in the range $0.11 < \theta_{\rm E}/{\rm mas} < 0.14$, 
are substantially smaller than the typical value. This suggests 
that either the mass of the lens is small $(M_{\rm tot} \ll 0.3~M_{\odot})$ 
or the lens is located at a large distance $(D_{\rm LS} \ll 1~{\rm kpc})$, 
where $D_{\rm LS} = D_{\rm S} - D_{\rm L}$ is the lens-source distance. 
Among two possibilities, i.e., small $M_{\rm tot}$ or large $D_{\rm L}$, 
the latter would be unlikely because the derived proper motions are 
in the range of $7.6 < \mu/({\rm mas}~{\rm yr}^{-1}) < 10.3$, suggesting 
that the lenses of both events are likely to be located in the Galactic disk.

\section{Physical Parameters}

In order to directly measure $M_{\rm tot}$ and $D_{\rm L}$, 
it is required to simultaneously detect $\theta_{\rm E}$ and 
the microlens parallax $\pi_{\rm E}$:
\begin{equation}
M_{\rm tot} = {\theta_{\rm E} \over \kappa\pi_{\rm E}};~~~~~D_{\rm L} = {{\rm AU} \over \pi_{\rm E}\theta_{\rm E} + \pi_{\rm S}},
\label{eq5}
\end{equation}
where $\pi_{\rm S} = {\rm AU}/D_{\rm S}$ is the parallax of the source. 
For both analyzed events, we measure the Einstein radii, but we cannot 
measure nor significantly constrain the microlens parallax signals 
due to the short timescales of the events. Hence, we cannot directly 
determine the physical lens parameters.

We can nevertheless constrain the physical parameters of the lenses from 
a Bayesian analysis based on the measured $t_{\rm E}$ and $\theta_{\rm E}$. 
For this, we first generate a large sample of lenses and sources, and 
distribute them over a model space using the Monte Carlo method. 
In this process, we adopt the Galactic model of \citet{jung18}. 
Next, we investigate the microlensing event rate of each lens-source pair, 
i.e., $\Gamma \propto \mu{\theta_{\rm E}}$. We then explore the posterior 
distributions of the primary mass $M_{1}$ and $D_{\rm L}$ by imposing 
$\Gamma$ and the measured $t_{{\rm E}, 1}$ and $\theta_{{\rm E}, 1}$ as a prior 
\footnote{$\theta_{{\rm E}, 1} = \theta_{\rm E} / \sqrt{1 + q}$ and $t_{{\rm E}, 1} = t_{\rm E} \theta_{{\rm E},1} / \theta_{\rm E}$.}.
Once $M_{1}$ and $D_{\rm L}$ are determined, we then estimate the secondary 
mass $M_{2}$ and the physical primary-secondary projected separation $a_{\bot}$ 
by 
\begin{equation}
M_{2} = qM_{1}
\label{eq6}
\end{equation}
and 
\begin{equation}
a_{\bot} = s{D_{\rm L}}{\theta_{\rm E}}, 
\label{eq6}
\end{equation}
respectively.

\begin{figure}[th]
\epsscale{0.9}
\plotone{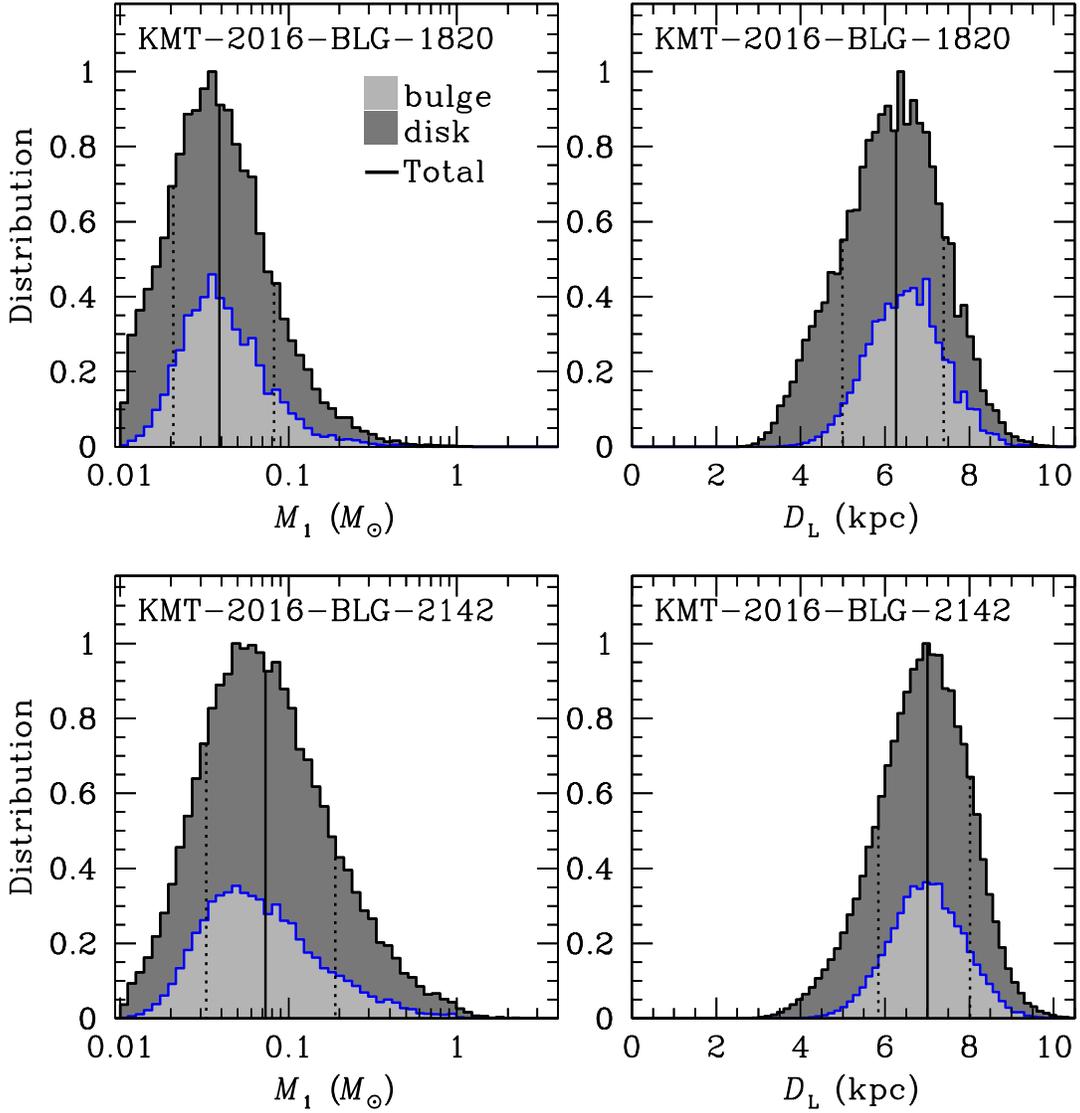}
\caption{\label{fig:six}
Posterior distributions of $M_{1}$ and $D_{\rm L}$ for KMT-2016-BLG-1820 
(upper left and right panels) and KMT-2016-BLG-2142 (lower left and right panels). 
In each panel, the total distribution (black line) is divided by bulge (grey) 
and disk (darkgrey) lenses. The vertical solid and dotted lines represent 
the median value and $68\%$ confidence intervals, respectively.     
}
\end{figure}

In Table~\ref{table:three}, we list the physical parameters of the 
individual events derived from our Bayesian analyses. 
The corresponding posterior distributions of $M_{1}$ (upper and lower left panels) 
and $D_{\rm L}$ (upper and lower right panels) are presented in Figure~\ref{fig:six}. 
We note that the physical values and their uncertainties are estimated based on the 
median values and $68\%$ confidence intervals of the distributions, respectively.

We find that the total lens masses of the individual events are, respectively, 
$M_{\rm tot} = 0.043_{-0.018}^{+0.043}~M_{\odot}$ and 
$M_{\rm tot} = 0.088_{-0.041}^{+0.120}~M_{\odot}$, well within the VLM regime. 
The binary lens of KMT-2016-BLG-1820 is composed of a brown dwarf-giant 
planet pair. For KMT-2016-BLG-2142, the binary host is a faint object whose mass 
is located near the hydrogen-burning limit ($\sim 0.075~M_{\odot}$; \citealt{burrows97}), 
while the companion is a dark object whose mass is located near the deuterium-burning 
limit ($\sim 13~M_{\rm J}$; \citealt{spiegel11}). The mass ratios and projected separations 
are $q = 0.113\pm0.003$ and $a_{\bot} = 1.08_{-0.24}^{+0.22}~{\rm AU}$ for KMT-2016-BLG-1820, 
and $q = 0.203\pm0.011$ and $a_{\bot} = 0.83_{-0.20}^{+0.15}~{\rm AU}$ for KMT-2016-BLG-2142. 
The lens distances of the individual events are $D_{\rm L} = 6.26_{-1.28}^{+1.14}$ kpc 
and $D_{\rm L} = 7.01_{-1.16}^{+1.01}$ kpc, respectively. Both lens systems are likely 
to be in the Galactic disk, consistent with the prediction based on the relatively high 
proper motions (see Table~\ref{table:two}). Therefore, the small $\theta_{\rm E}$ values 
in both events originate from the small lens masses combined with the lens locations 
being closer to the lensed stars than to Earth.

\section{Discussion}

We found two possible planetary mass companions around very low-mass hosts 
from the analysis of two microlensing events. In both events, the lensing 
perturbations were clearly captured and densely covered by the KMTNet survey. 
These prove the capability of KMTNet experiment to identify even brief 
anomalies in very short-timescale $(t_{\rm E} \sim 5~{\rm day})$ events.

\begin{figure}[th]
\epsscale{0.9}
\plotone{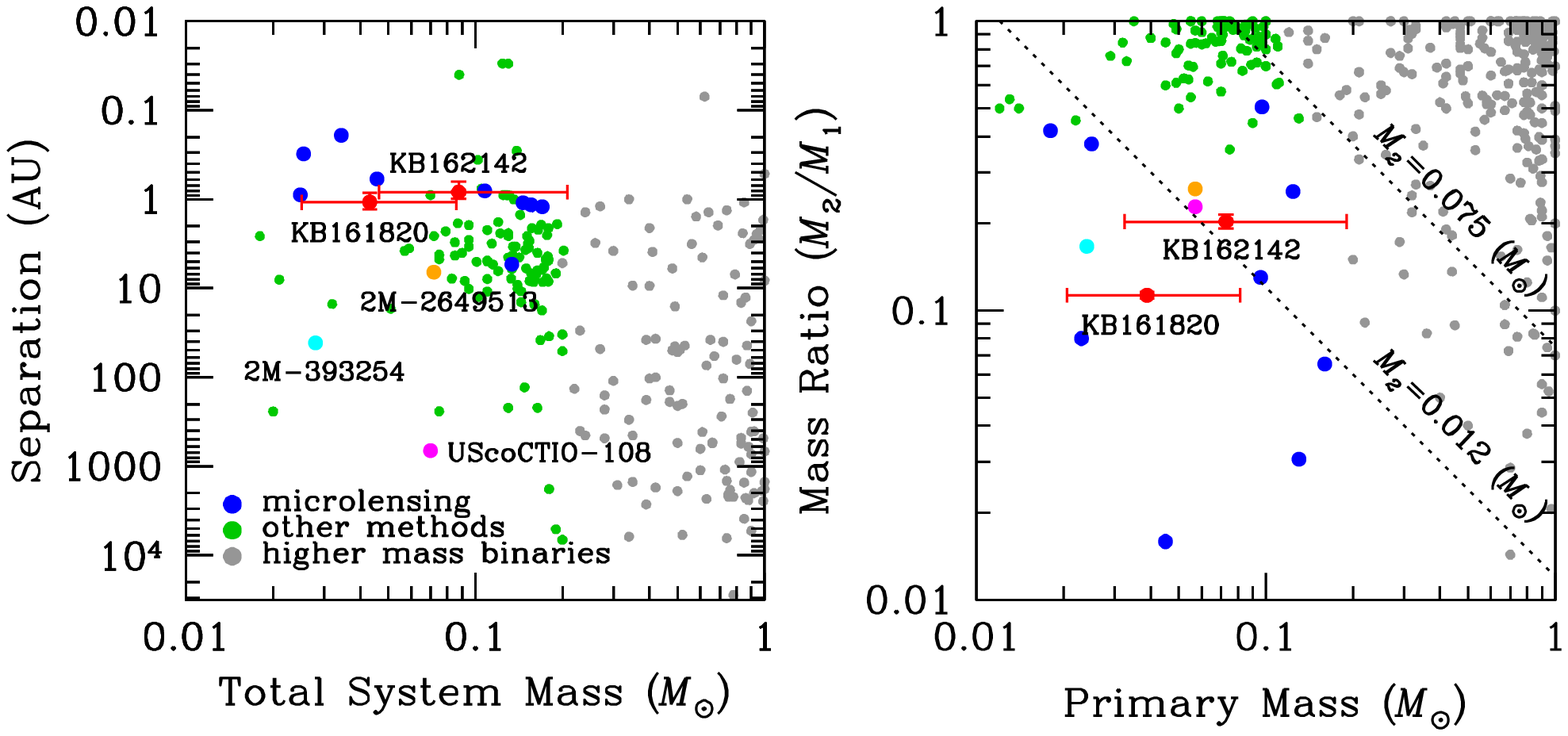}
\caption{\label{fig:seven}
Total mass vs. separation (left panel) and primary mass vs. mass ratio (right panel) 
distributions for known binaries. The two red dots are the binaries 
reported in this work. The blue dots are microlensing binaries for which the 
projected separations are unambiguously determined and the total masses are less 
than $0.2~M_{\odot}$ \citep{choi13, han13, han17, street13, park13, jung15, jung18}. 
The green dots are VLM binaries discovered by other methods, while the grey dots 
are some higher mass binaries. The yellow, magenta, and cyan dots are the three binaries 
that have similar mass ratios and total masses to the two reported binaries. 
The values are obtained from the Very-Low-Mass Binaries Archive (http://www.vlmbinaries.org) and other references 
\citep{basri99,lane01,allers09,allers10,faherty11,burgasser08,burgasser11,
burgasser12,burgasser16,dhital11,liu11,liu12,duchene13,luhman13,sahlmann13,dupuy16,dupuy17}. 
For given primary masses ($M_{1}$) and mass ratios ($q$), companion masses ($M_{2}$) corresponding to 
star/brown-dwarf and brown-dwarf/planet boundaries are presented by two dotted lines. 
}
\end{figure}

In Figure~\ref{fig:seven}, we compare the physical properties of these two lens systems 
to those of previously known VLM ($M_{\rm tot} \leq 0.2 M_{\odot}$) binaries and some higher mass binaries. The figure 
clearly shows that the fraction of microlensing binaries are high in the low-mass-ratio 
($q \leq 0.5$) and close-separation ($a_{\bot} \leq 1~{\rm AU}$) regions. Among known VLM 
binaries discovered from other methods, we find that three binaries have similar mass ratios 
and total masses to our results: 
2MASS J13153094-2649513 \citep{burgasser11} with $(q, M_{\rm tot}/M_{\odot}) \sim (0.27, 0.073)$, 
UScoCTIO-108 \citep{bejar08} with $(q, M_{\rm tot}/M_{\odot}) \sim (0.20, 0.074)$, 
and 2MASS J1207334-393254 \citep{chaubin04} with $(q, M_{tot}/M_{\odot}) \sim (0.20, 0.028)$. 
However, these binaries have wide separations ($> 6~{\rm AU}$) and were discovered in young 
associations, implying that the sensitivity regimes of microlensing and other detection method 
are quite different. Hence, these prove that microlensing method can help to achieve 
the unbiased VLM binary samples by complementing other methods.

The distinctive sensitivity regimes of microlensing make it possible to improve our 
understanding of VLM formation mechanisms. From many observational works, it has been 
suggested that the binary properties of VLM objects are quite different from those of 
higher-mass stars (e.g., \citealt{burgasser07}). For example, about $15 \sim 20$ per cent 
of VLM systems form as binaries having roughly equal mass ratios with peak separations of 
$\sim 3~{\rm AU}$, in contrast to their stellar counterparts for which the binarity 
frequencies are about $30 \sim 60$ per cent and the mass ratios and separations are 
broadly spread \citep{duquennoy91,fischer92}. Until now, these statistical differences 
have been widely used as evidence for distinct formation scenarios between VLM objects 
and stars. However, the statistical properties of VLM binaries are largely based on the 
samples collected by direct imaging, implying that the distributions are strongly affected 
by selection effects and detection biases. In fact, we still do not know clearly whether 
there exists a significant number of VLM binaries in low-mass-ratio $(q \leq 0.5)$ and 
close-separation $(\leq 3~{\rm AU})$ regions, although it has been suggested that these 
VLM binaries are as frequent as their counterparts (i.e., high mass ratios and wide 
separations). Furthermore, understanding the fraction of such VLM binaries is very important, 
because it provides a key constraint for the question that whether VLM objects form in a 
similar manner to hydrogen-burning stars or whether they require additional (or different) 
formation processes. As shown in Figure~\ref{fig:seven}, microlensing can enrich VLM binary 
samples in these regimes. Hence, the microlensing method can play an important role in 
providing the empirical constraints for exploring the origins of VLM objects.


\acknowledgments
This research has made use of the KMTNet system operated by the Korea 
Astronomy and Space Science Institute (KASI) and the data were obtained at 
three host sites of CTIO in Chile, SAAO in South Africa, and SSO in Australia. 
C. Han was supported by grant 2017R1A4A1015178 of the National Research 
Foundation of Korea. Work by WZ and AG were supported by AST-1516842 from 
the US NSF. WZ and AG were supported by JPL grant 1500811. 
AG is supported from KASI grant 2016-1-832-01.

\begin{deluxetable}{lrrr}
\tablecaption{Lensing Parameters\label{table:one}}
\tablewidth{0pt}
\tablehead{
\multicolumn{1}{l}{Parameters} &
\multicolumn{1}{c}{KMT-2016-BLG-1820} &
\multicolumn{2}{c}{KMT-2016-BLG-2142} \\
\multicolumn{1}{l}{} &
\multicolumn{1}{r}{} &
\multicolumn{1}{c}{Close} &
\multicolumn{1}{c}{Wide} 
}
\startdata
$\chi^2$/dof            &     7510.0/7497      &     6457.1/6444      &     6498.2/6444        \\
$t_{0}$ (${\rm HJD'}$)  & 7632.19 $\pm$ 0.02   &  7612.54 $\pm$ 0.04  &    7612.25 $\pm$ 0.06  \\
$u_{0}$                 &    0.24 $\pm$ 0.01   &     0.14 $\pm$ 0.01  &       0.15 $\pm$ 0.01  \\
$t_{\rm E}$ (days)      &    4.81 $\pm$ 0.03   &     5.15 $\pm$ 0.22  &       6.08 $\pm$ 0.25  \\
$s$                     &    1.40 $\pm$ 0.01   &     0.97 $\pm$ 0.01  &       1.21 $\pm$ 0.02  \\
$q$ ($10^{-1}$)         &    1.13 $\pm$ 0.03   &     2.03 $\pm$ 0.11  &       2.09 $\pm$ 0.11  \\
$\alpha$ (rad)          &    3.04 $\pm$ 0.01   &     3.96 $\pm$ 0.03  &       3.94 $\pm$ 0.03  \\
$\rho_\ast$ ($10^{-3}$) &    6.56 $\pm$ 0.21   &     6.42 $\pm$ 0.34  &       5.09 $\pm$ 0.21                  
\enddata 
\vspace{0.05cm}
\tablecomments{
${\rm HJD}'= {\rm HJD}-2450000~{\rm days}$
}
\end{deluxetable}

\begin{deluxetable}{lrr}
\tablecaption{Source Star Properties\label{table:two}}
\tablewidth{0pt}
\tablehead{
\multicolumn{1}{l}{Parameters} &
\multicolumn{1}{r}{KMT-2016-BLG-1820} &
\multicolumn{1}{r}{KMT-2016-BLG-2142} 
}
\startdata
$\Delta(V-I, I)$                  &    $(-0.42\pm0.03, 3.37\pm0.02)$  &    $(-0.23\pm0.04, 3.90\pm0.02)$ \\ 
$(V-I, I)_{0,\rm GC}$             &    $(1.06\pm0.07, 14.42\pm0.09)$  &    $(1.06\pm0.07, 14.43\pm0.09)$ \\
$(V-I, I)_{0}$                    &    $(0.64\pm0.08, 17.79\pm0.09)$  &    $(0.83\pm0.08, 18.33\pm0.09)$ \\
$\theta_{*}$ $({\mu}{\rm as})$    &    $0.807\pm0.072$                &    $0.781\pm0.072$               \\
$\theta_{\rm E}$ (mas)            &    $0.123\pm0.012$                &    $0.122\pm0.013$               \\     
$\mu$ $({\rm mas}~{\rm yr}^{-1})$ &    $9.341\pm0.882$                &    $8.622\pm0.934$                    
\enddata 
\vspace{0.05cm}
\end{deluxetable}

\begin{deluxetable}{lrr}
\tablecaption{Physical Parameters\label{table:three}}
\tablewidth{0pt}
\tablehead{
\multicolumn{1}{l}{Parameters} &
\multicolumn{1}{c}{KMT-2016-BLG-1820} &
\multicolumn{1}{c}{KMT-2016-BLG-2142} 
}
\startdata
$M_{1}$ $(M_{\odot})$             &    $0.039_{-0.018}^{+0.043}$  &  $0.073_{-0.040}^{+0.117}$   \\ 
$M_{2}$ $(M_{\rm J})$             &    $4.57_{-2.14}^{+5.03}$     &  $15.49_{-8.58}^{+24.99}$    \\ 
$D_{\rm L}$ (kpc)                 &    $6.26_{-1.28}^{+1.14}$     &  $7.01_{-1.16}^{+1.01}$      \\ 
$a_{\bot}$ (AU)                   &    $1.08_{-0.24}^{+0.22}$     &  $0.83_{-0.20}^{+0.15}$      \\ 
\enddata 
\vspace{0.05cm}
\end{deluxetable}

\begin{thebibliography}{99}

\bibitem[Alard \& Lupton(1998)]{alard98}
Alard, C., \& Lupton, Robert H. 1998, \apj, 503, 325

\bibitem[Albrow et al.(2009)]{albrow09}
Albrow, M. D., Horne, K., Bramich, D. M., et al. 2009, \mnras, 397, 2099

\bibitem[Allers et al.(2010)]{allers10}
Allers, K. N., Liu, M. C., Dupuy, T. J., \& Cushing, M. C. 2010, \apj, 715, 561

\bibitem[Allers et al.(2009)]{allers09}
Allers, K. N., Liu, M. C., Shkolnik, E., et al. 2009, \apj, 697, 824

\bibitem[Bardalez Gagliuffi et al.(2014)]{bardalez14}
Bardalez Gagliuffi, D. C., Burgasser, A. J., Gelino, C. R., et al. 2014, \apj, 794, 143. 

\bibitem[Basri \& Mart\'in (1999)]{basri99}
Basri, G., \& Mart\'in, E. L. 1999, \aj, 118, 2460

\bibitem[Bate(2009)]{bate09}
Bate, M. R. 2009, \mnras, 397, 232.

\bibitem[Bensby et al.(2013)]{bensby13}
Bensby, T., Yee, J. C., Feltzing, S., et al. 2013, \aap, 549, 147

\bibitem[Bessell \& Brett(1988)]{bessell88}
Bessell, M. S., \& Brett, J. M. 1988, \pasp, 100, 1134

\bibitem[B\'{e}jar et al.(2008)]{bejar08}
B\'{e}jar, V. J. S., Zapatero Osorio, M. R., P\'{e}rez Garrido, A., et al. 2008, \apj, 673, 185

\bibitem[Burgasser et al.(2016)]{burgasser16}
Burgasser, A. J., Blake, C. H., Gelino, C. R., et al. 2016, \apj, 827, 25

\bibitem[Burgasser et al.(2008)]{burgasser08}
Burgasser, A. J., Liu, M. C., Ireland, M. J., et al. 2008, \apj, 681, 579

\bibitem[Burgasser et al.(2012)]{burgasser12}
Burgasser, A. J., Luk, C., Dhital, S., et al. 2012, \apj, 757, 110

\bibitem[Burgasser et al.(2007)]{burgasser07}
Burgasser, A. J., Reid, I. N., Siegler, N., et al. 2007, in Protostars and Planets V, ed. B. Reipurth, D. Jewitt, \& K. Keil (Tucson, AZ: Univ. Arizona Press), 427

\bibitem[Burgasser et al.(2011)]{burgasser11}
Burgasser, A. J., Sitarski, B. N., Gelino, C. R., et al. 2011, \apj, 739, 49

\bibitem[Burrows et al.(1997)]{burrows97}
Burrows, A., Marley, M., Hubbard, W. B., et al. 1997, \apj, 491, 856

\bibitem[Chauvin et al.(2004)]{chaubin04}
Chauvin, G., Lagrange, A.-M., Dumas, C., et al. 2004, \aap, 425, L29

\bibitem[Choi et al.(2013)]{choi13}
Choi, J.-Y., Han, C., Udalski, A., et al. 2013, \apj, 768, 129

\bibitem[Chung et al.(2017)]{chung17}
Chung, S.-J., Zhu, W., Udalski, A., et al. 2017, \apj, 838, 154

\bibitem[Close et al.(2007)]{close07}
Close, L. M., Zuckerman, B., Song, I., et al. 2007, \apj, 660, 1492

\bibitem[Dhital et al.(2011)]{dhital11}
Dhital, S., Burgasser, A. J., Looper, D. L., \& Stassun, K. G. 2011, \aj, 141, 7

\bibitem[Dominik(1998)]{dominik98}
Dominik, M. 1998, \aap, 329, 361

\bibitem[Dominik(1999)]{dominik99}
Dominik, M. 1999, \aap, 349, 108

\bibitem[Duch\^{e}ne et al.(2013)]{duchene13}
Duch\^{e}ne, G., Bouvier, J., Moraux, E., et al. 2013, \aap, 555, A137

\bibitem[Dupuy et al.(2016)]{dupuy16}
Dupuy, T. J., Forbrich, J., Rizzuto, A., et al. 2016, \apj, 827, 23

\bibitem[Dupuy \& Liu(2017)]{dupuy17}
Dupuy T. J., \& Liu M. C., 2017, \apjs, 231, 15

\bibitem[Duquennoy \& Mayor,(1991)]{duquennoy91}
Duquennoy, A., \& Mayor, M. 1991, \aap, 248, 485

\bibitem[Faherty et al.(2011)]{faherty11}
Faherty, J. K., Burgasser, A. J., Bochanski, J. J., et al. 2011, \aap, 141, 71

\bibitem[Fischer \& Marcy(1992)]{fischer92}
Fischer, D. A., \& Marcy, G. W. 1992, \apj, 396, 178

\bibitem[Gould(1992)]{gould92}
Gould, A.\ 1992, \apj, 392, 442

\bibitem[Gould(2004)]{gould04}
Gould, A. 2004, \apj, 606, 319

\bibitem[Gould et al.(2009)]{gould09}
Gould, A., Udalski, A., Monard, B., et al. 2009, \apjl, 698, L147

\bibitem[Griest \& Safizadeh(2012)]{griest98}
Griest, K., \& Safizadeh, N. 1998, \apj, 500, 37

\bibitem[Han et al.(2013)]{han13}
Han, C., Jung, Y. K., Udalski, A., et al. 2013, \apj, 778, 38


\bibitem[Han et al.(2017)]{han17}
Han, C., Udalski, A., Sumi, T., et al. 2017, \apj, 843, 59

\bibitem[Jung et al.(2013)]{jung13}
Jung, Y. K., Han, C., Gould, A., \& Maoz, D. 2013, \apjl, 768, L7J

\bibitem[Jung et al.(2018)]{jung18}
Jung, Y. K., Udalski, A., Gould, A., et al. 2018, \aj, 155, 219

\bibitem[Jung et al.(2015)]{jung15}
Jung, Y. K., Udalski, A., Sumi, T., et al. 2015, \apj, 798, 123

\bibitem[Kervella et al.(2004)]{kervella04}
Kervella P., Th\'{e}venin F., Di Folco E., S\'{e}gransan D., 2004, \aap, 426, 297

\bibitem[Kim et al.(2018a)]{kim18a}
Kim, D.-J., Kim, H.-W., Hwang, K.-H., et al., 2018a, \aj, 155, 76

\bibitem[Kim et al.(2018b)]{kim18b}
Kim, H.-W.., Hwang, K.-H., Kim, D.-J., et al., 2018b, AAS submitted, arXiv:1804.03352 

\bibitem[Kim et al.(2016)]{kim16}
Kim, S.-L., Lee, C.-U., Park, B.-G., et al. 2016, JKAS, 49, 37

\bibitem[Lane et al.(2001)]{lane01}
Lane, B. F., Zapatero Osorio, M. R., Britton, M. C., et al. 2001, \apj, 560, 390

\bibitem[Liu et al.(2011)]{liu11}
Liu, M. C., Delorme, P., Dupuy, T. J., et al. 2011, \apj, 740, 108

\bibitem[Liu et al.(2012)]{liu12}
Liu, M. C., Dupuy, T. J., Bowler,B. P., et al. 2012, \apj, 758, 57

\bibitem[Luhman(2013)]{luhman13}
Luhman, K. L. 2013, \apj, 767, L1

\bibitem[Nakajima et al.(1995)]{nakajima95}
Nakajima, T., Oppenheimer, B. R., Kulkarni, S. R., et al. 1995, Nat, 378, 463

\bibitem[Nataf et al.(2013)]{nataf13}
Nataf, D. M., Gould, A., Fouqu\'{e}, P., et al. 2013, \apj, 769, 88

\bibitem[Padoan \& Nordlund(2004)]{padoan04}
Padoan, P., \& Nordlund, A. 2004, \apj, 617, 559

\bibitem[Park et al.(2013)]{park13}
Park, H., Udalski, A., Han, C., et al. 2013, \apj, 778, 134

\bibitem[Rebolo et al.(1995)]{rebolo95}
Rebolo, R., Zapatero Osorio, M. R., \& Mart\'{i}n, E. L. 1995, Nat, 377, 129

\bibitem[Reipurth \& Clarke(2001)]{reipurth01}
Reipurth, B., \& Clarke, C. 2001, \aj, 122, 432

\bibitem[Sahlmann et al.(2013)]{sahlmann13}
Sahlmann, J., Lazorenko, P. F., S\'{e}gransan, D., et al. 2013, \aap, 556, 133

\bibitem[Spiegel et al.(2011)]{spiegel11}
Spiegel, D. S., Burrows, A., \& Milsom, J. A. 2011, \apj, 727, 57

\bibitem[Stamatellos \& Whitworth(2004)]{stamatellos09}
Stamatellos, D., \& Whitworth, A. P. 2009, \mnras, 392, 413

\bibitem[Street et al.(2013)]{street13}
Street, R., Choi, J.-Y., Tsapras, Y., et al. 2013, \apj, 763, 67

\bibitem[Szyma\'{n}ski et al.(2011)]{szymanski11}
Szyma\'{n}ski, M. K., Udalski, A., Soszy\'{n}ski, I., et al. 2011, AcA, 61, 83

\bibitem[Whitworth \& Zinnecker(2004)]{whitworth04}
Whitworth A. P., \& Zinnecker H. 2004, \aap, 427, 299

\bibitem[Wo\'{z}niak(2000)]{wozniak00}
Wo\'{z}niak, P. R. 2000, Acta Astron., 50, 42

\bibitem[Yee et al.(2012)]{yee12}
Yee, J. C., Shvartzvald, Y., Gal-Yam, A., et al. 2012, \apj, 755, 102

\bibitem[Yoo et al.(2004)]{yoo04}
Yoo, J., DePoy, D. L., Gal-Yam, A., et al. 2004, \apj, 603, 139

\bibitem[Zhu et al.(2016)]{zhu16}
Zhu, W., Calchi Novati, S., Gould, A., et al. 2016, \apj, 825, 60

\end{thebibliography}
\end{document}